\documentclass[namedreferences]{kluwer}
\newdisplay{guess}{Conjecture}

\begin{document}                                                                                   
\begin{article}
\begin{opening}         
\title{Epicyclic frequencies derived from the effective potential:
\\simple and practical formulae} 
\author{M.A. \surname{Abramowicz}\footnote{On leave from Department of Astrophysics, G{\"o}teborg University. Address: Department of Theoretical Physics, Chalmers University, S-412-96 G{\"o}teborg, Sweden, {\tt marek@fy.chalmers.se} }}
\author{W. \surname{Klu{\'z}niak}\footnote{Copernicus Astronomical Centre, Warsaw \& Institute of Astronomy, Zielona G{\'o}ra University. Address: Copernicus Centre, Bartycka 18, PL-00-716 Warszawa, Poland, {\tt wlodek@camk.edu.pl }}} 
\institute{Nordita, Copenhagen, Denmark}

\author{Marek A. \surname{Abramowicz}}  
\runningauthor{M.A.~Abramowicz \& W.~Klu{\'z}niak}
\runningtitle{Epicyclic frequencies}

%\date{November 1, 2004}

\begin{abstract}
We present and discuss a short and simple derivation of orbital epicyclic frequencies for circular geodesic orbits in stationary and axially symmetric spacetimes. Such spacetimes include as special cases analytically known black hole Kerr and Schwarzschild spacetimes, as well as the analytic Hartle-Thorne spacetime and all numerically constructed spacetimes relevant for rotating neutron stars. Our derivation follows directly from energy and angular momentum conservation and it uses the concept of the effective potential. It has never been published, except for a few special cases, but it has already become a part of the common knowledge in the field. 
\end{abstract}

\end{opening}           

%%%%%%%%%%%%%%%%%%%%%%%%%%%%%%%%%%%%%%%%%%%%%%%%%%%%%%%%%%%%%%%%%%%%
%%
\section{Introduction} 
\label{Intro-1}
%%
%%%%%%%%%%%%%%%%%%%%%%%%%%%%%%%%%%%%%%%%%%%%%%%%%%%%%%%%%%%%%%%%%%%%

Properties of congruences of nearly circular geodesic orbits in stationary and axially symmetric spacetimes are studied because of their fundamental role in the theory of accretion disks around compact objects with strong gravity. The Keplerian angular frequency $\Omega_K$, the radial epicyclic frequency $\omega_R$ and the vertical epicyclic frequency $\omega_V$ are the most important characteristics of these orbits. Analytic formulae for the three frequencies in Schwarzschild, Kerr and Hartle-Thorne metrics have been published many times by several authors
(see e.g. \opencite{Wal1984}; \opencite{OkaKatFuk1987}; \opencite{Perez97}; \opencite{NowLeh1999}; \opencite{AbrAlmKluTha2003}) and are well-known. 

Here, we recall our unpublished \cite{AbraKlu2000} derivation of the epicyclic frequencies in the general case. This fully relativistic derivation is remarkably simple: just four short lines of very transparent, easy to check algebra. It uses only invariantly defined quantities that have obvious physical meaning, and it closely follows the standard Newtonian derivation that is equally short and based on the concept of the effective potential.\footnote{The same simple method was used in the text book by \citeauthor{Wal1984} \shortcite{Wal1984} in the special case of a static, spherically symmetric metric. The general formulae derived by \inlinecite{Mar2000} are complicated and physically unclear---it is not easy to find and correct the few missprints that they contain.} 

The standard Newtonian derivation of epicyclic frequencies is briefly recalled in section \ref{Newton-2}, where we also make a few comments on these properties of the orbital frequencies that are not commonly appreciated. In section \ref{formalism-3}, for the completeness of presentation, we review some basic facts concerning relativistic formalism that are directly relevant to the derivation of the orbital frequencies in Einstein's theory. The reader who is interested only in our four-line derivation may skip the next two sections and start reading from section \ref{derivation-4}. 

%%%%%%%%%%%%%%%%%%%%%%%%%%%%%%%%%%%%%%%%%%%%%%%%%%%%%%%%%%%%%%%%%%%%
%%
\section{Nearly circular geodesic motion: Newton's theory}
\label{Newton-2}
%%
%%%%%%%%%%%%%%%%%%%%%%%%%%%%%%%%%%%%%%%%%%%%%%%%%%%%%%%%%%%%%%%%%%%%

In Newton's theory, circular geodesic orbits (i.e., Keplerian circular orbits of free particles) occur at locations $r=r_0, \,\,\,\theta = \theta_0$, where the effective potential ${U}_{\rm eff}$ has an extremum with respect to both radial $r$ and polar-angle $\theta$ coordinates\footnote{We use the standard spherical coordinates  $[t, r, \theta, \phi]$.},

\begin{equation}
\label{Newton-effective-potential}
\left ( {{\partial { U}_{\rm eff}} \over {\partial r}} \right) = 0 = \left ( {{\partial { U}_{\rm eff}} \over {\partial \theta}} \right),~~{ U}_{\rm eff} \equiv \Phi (r, \theta) + {{ L}^2 \over {2r^2\sin^2 \theta}}. 
\end{equation}

\noindent The derivatives are taken at a constant angular momentum ${ L}$. The gravitational potential $\Phi (r, \theta)$ is assumed to be stationary, $\partial \Phi/\partial t =0$, axially symmetric $\partial \Phi/\partial \phi =0$, and to have the equatorial symmetry plane at $\theta = \pi/2$. From the last assumption it follows that one can take $\theta_0 = \pi/2$, i.e., that there are circular orbits located in the equatorial plane.

Note, that from the solution ${ L} = { L}_K(r)$ of equation (\ref{Newton-effective-potential}) it follows immediately that in the equatorial plane the radial distributions of Keplerian angular momentum ${ L}_K$ and Keplerian angular velocity $\Omega_K$ are given by\footnote{Obviously, Keplerian orbits are impossible in the region where $(\partial \Phi /\partial r) < 0$.},

\begin{equation}
\label{Newton-Keplerian}
{ L}_K = \pm \left (r^3\, {{\partial \Phi} \over {\partial r}}\right)^{1\over 2}, ~~ \Omega_K = {{{ L}_K} \over r^2} = \pm \left (r^{-1}\, {{\partial \Phi} \over {\partial r}}\right)^{1\over 2}.
\end{equation}

\noindent Here, the derivatives are taken at the equatorial plane, $\theta = \pi/2$.

In the linear regime we can consider separately the case of purely radial, $\delta r = r -r_0$, $\delta \theta = 0$, and purely vertical, $\delta \theta = \theta - \theta_0$, $\delta r =0$, epicyclic oscillations 
about circular orbits in the equatorial plane.
Take a ``perturbed'' Keplerian orbit which may be slightly not circular,
or slightly off equatorial plane, and has 
the same angular momentum as the original circular orbit ${ L} = { L}_0$, but with a slightly larger energy ${ E} = { E}_0 + \delta { E}$. Because circular orbits occur at extrema of the effective potential, the Taylor expansion of ${ U}_{\rm eff}$ near a circular orbit starts from the second-order term,

\begin{equation}
\label{Newton-effective-1}
\delta { U}_{\rm eff} =  
{1\over 2} \left ( {{\partial^2 { U}_{\rm eff}} \over {\partial r^2}} \right) \delta r^2,~~
%\end{equation}
%
%\begin{equation}
%\label{Newton-effective-2}
\delta { U}_{\rm eff} =  
{1\over 2} \left ( {{\partial^2 { U}_{\rm eff}} \over {\partial \theta^2}} \right) \delta \theta^2.
\end{equation}

\noindent  A small change in the orbital energy $\delta { E}$ may be therefore expressed as,

\begin{equation}
\label{Newton-perturbed-11}
\delta { E} = {1\over 2} \left ( {{\partial^2 { U}_{\rm eff}} \over {\partial r^2}} \right) \delta r^2 + {1\over 2} (\delta {\dot r})^2,~~
%\end{equation}
%
%\begin{equation}
%\label{Newton-perturbed-12}
\delta { E} = {1\over 2} \left ( {{\partial^2 { U}_{\rm eff}} \over {\partial \theta^2}} \right) \delta \theta^2 + {1\over 2} r^2\,(\delta {\dot \theta})^2.
\end{equation}

\noindent The energy of the perturbed orbit is obviously a constant of motion, and thus $\delta {\dot { E}} = 0$, from which it follows that, 

\begin{equation}
\label{Newton-perturbed-21}
0 = \delta {\dot r} \left [ \left ( {{\partial^2 { U}_{\rm eff}} \over {\partial r^2}} \right) \delta r + \delta {\ddot r} \right ],~~
0 = \delta {\dot \theta} \left [ \left ( {{\partial^2 { U}_{\rm eff}} \over {\partial \theta^2}} \right) \delta \theta + r^2\, \delta {\ddot \theta} \right ].
\end{equation}

\noindent We are interested in a non-trivial solution $\delta {\dot r} \not = 0 \not = \delta {\dot \theta}$, for which the last two equations take the well-known form of a simple harmonic oscillator,

\begin{equation}
\label{Newton-oscillator}
0 = \omega_R^{~2}\delta r + \delta {\ddot r},~~0=\omega_V^{~2}\delta \theta + \delta {\ddot \theta}
\end{equation}

\noindent with the squared eigenfrequencies being the second derivatives of the effective potential with respect to proper geodesic distances in radial $R$ and vertical $V$ directions. The radial and vertical directions are defined in a {\it coordinate independent} way that for the particular case of spherical coordinates yields  $dR = dr$, $dV = r d\theta$. Thus, one may write,

\begin{equation}
\label{Newton-eigenfrequency}
\omega_R^{~2} \equiv \left ( {{\partial^2 { U}_{\rm eff}} \over {\partial R^2}} \right)= \left ( {{\partial^2 { U}_{\rm eff}} \over {\partial r^2}} \right), ~~
\omega_V^{~2} \equiv \left ( {{\partial^2 { U}_{\rm eff}} \over {\partial V^2}} \right) = {1\over r^2} \left ( {{\partial^2 { U}_{\rm eff}} \over {\partial \theta^2}} \right).
\end{equation}

\noindent From equation (\ref{Newton-eigenfrequency}), and  equations (\ref{Newton-effective-potential}), (\ref{Newton-Keplerian}) it follows that,

\begin{equation}
\label{Newton-Rayleigh}
\omega_R^{~2} = \Omega_K^2\,\left ( {{d \ln{ L}_K^2} \over {d \ln r}} \right),~~
\omega_V^{~2} = \Omega^2_K + {1\over r^2}\left ( {{\partial^2 \Phi} \over {\partial \theta^2}} \right ) .
\end{equation}

\noindent Several important deductions could be made from (\ref{Newton-Rayleigh}). Firstly, it is obvious that $\omega_R = \Omega_K$ everywhere if and only if ${ L}_K^2 \propto r$, and this together with equation (\ref{Newton-Keplerian}) implies $\Phi \propto 1/r$. Secondly, $\omega_V = \Omega_K$ if (but not only if) the gravitational potential is spherically symmetric. Therefore, one concludes that in the special but important case of the $\Phi = -GM/r$ potential, all three characteristic orbital frequencies are equal and their squares are positive, 

\begin{equation}
\label{Newton-point-mass}
\omega_R^{~2}  = \omega_V^{~2} = \Omega_K^{~2} = {GM\over r^3} > 0.
\end{equation}

\noindent Here $M$ is the mass of the gravity source. The equality of the three frequencies means that the orbits in the $\Phi = -GM/r$ potential are periodic and closed (a fact known already to Kepler, who discovered that they are ellipses). The positive squares imply stability of both radial and vertical oscillations, i.e., the dynamical stability of circular orbits in the $\Phi = -GM/r$ potential. 

The third important deduction from (\ref{Newton-Rayleigh}) is that in the general case, with a potential $\Phi(r,\theta) \not = -GM/r$, the radial epicyclic oscillations around Keplerian circular orbits are stable (i.e., $\omega_R^{~2} > 0$) if and only if the Keplerian angular momentum, $|{ L}_K(r)|$, is an increasing function of the radius $r$ of these orbits. This statement, of course, is just a special case of the well-known Rayleigh stability criterion. 

The last deduction concerns the stability of the vertical epicyclic oscillations. If a non-spherical gravitational potential $\Phi(r,\theta)$ has a {\it positive} second derivative with respect to $\theta$ at the equatorial symmetry plane, the vertical oscillations are stable. Only if the second derivative is negative and sufficiently large, these oscillations may be unstable.

Let us illustrate the above general discussion in terms of a specific example. The gravitational potential expansion in terms of spherical harmonics (with $m=0$ because of the axial symmetry) yields,

\begin{equation}
\label{harmonics}
\Phi(r,\theta) = - {GM \over r} - {{G Q P_2  (\cos \theta)} \over r^3} + 
..., ~~P_2(\cos \theta) \equiv {1\over 2}(3\cos^2 \theta - 1).
\end{equation}

\noindent The higher order terms are ${\cal O}(1/r^5)$ because of the equatorial plane symmetry. The quadrupole moment $Q$ is {\it negative} for an oblate mass distribution, and because rotation typically produces a bulge at the equator, $Q \le 0$ is the realistic case to consider. Assuming this, and neglecting higher multipoles, we deduce from (\ref{harmonics}) that,

\begin{equation}
\label{quadrupole-vertical-stability}
{1\over r^2}\left ( {{\partial^2 \Phi} \over {\partial \theta^2}} \right ) = 
- {{ 3 G Q} \over r^5} \ge 0,
\end{equation}

\begin{equation}
\label{quadrupole-radial-stability}
L_K^2 = G M r - {3 \over 2} {{G Q} \over r} > 0,~~
\left ( {{d \ln{ L}_K^2} \over {d \ln r}} \right) =
{{2 M r^2 + 3 Q} \over {2 M r^2 - 3 Q}}
\end{equation}

\noindent We see from (\ref{quadrupole-vertical-stability}) that a negative quadrupole moment never destabilizes vertical oscillations. Equation (\ref{quadrupole-radial-stability}) seems to indicate that a negative quadrupole moment destabilizes radial oscillations at small enough radii\footnote{If the quadrupole is positive, then for $r < r_Q$ no circular orbits are possible. This provides an example to a situation mentioned in the footnote 3.}

\begin{equation}
\label{quadrupole-destabilization}
r < r_Q \equiv  \sqrt{ {{3|Q|}\over {2M}}}. 
\end{equation}

\noindent However, let us write $|Q| = q M R^2_*$, with $R_*$ being the equatorial radius of the star and $q$ being a dimensionless parameter. A body with fixed $M$ and $R_*$ will have the maximal quadrupole if the whole mass is placed in an infinitesimally thin ring of matter located at $r = R_*,~\theta = \pi/2$. It is trivial to calculate the quadrupole moment in this case,

\begin{subequation}
\label{quadrupole}
\begin{eqnarray}
Q &\equiv& \int_0^M r^2 P_2 (\cos \theta) dM\\ 
&=& \left [ r^2 P_2 (\cos \theta) \right ]_{r = R_*,\theta=\pi/2}
\int_0^M dM
= - {1\over 2} R_*^2 M.
\end{eqnarray}
\end{subequation}

\noindent The above formula proves that the destabilization condition (\ref{quadrupole-destabilization}) is impossible to fulfill outside the star, because it implies $r < (\sqrt{3}/2)R_*<R_*$. 

One concludes that according to Newton's theory, the quadrupole moment cannot destabilize circular orbits (in the equatorial plane) of particles orbiting a body. In consciousness of many astrophysicists this conclusion constituteses a ``proof'' that circular orbits around Newtonian bodies, are {\it always} stable. Thus, the recent finding of \inlinecite{AmsEtAL2002} that near very ra\-pid\-ly ro\-tating Newtonian Maclaurin spheroids there {\it are} unstable orbits (with $\omega^2_R < 0$) came as a  surprise.  \inlinecite{KlEtAL2001} found that radial epicyclic oscillations are destabilised by {\it octupole} and higher moments in the harmonic expansion of the potentials.

%%%%%%%%%%%%%%%%%%%%%%%%%%%%%%%%%%%%%%%%%%%%%%%%%%%%%%%%%%%%%%%%%%%%
%%
\section{Stationary, axially symmetric spacetimes}
\label{formalism-3}
%%
%%%%%%%%%%%%%%%%%%%%%%%%%%%%%%%%%%%%%%%%%%%%%%%%%%%%%%%%%%%%%%%%%%%%

 We assume\footnote{We use everywhere the $(+\,-\,-\,-)$ signature and, occasionally, $c = 1 = G$ units.} that spacetimes considered here are stationary and axially symmetric, which means that they admit two Killing vectors $\eta^i$ and $\xi^i$, which obey,

\begin{equation}
\label{Killing}
\nabla_{(i}\eta_{k)} = 0,~~\nabla_{(i}\xi_{k)} = 0,~~ \xi^k \nabla_k \eta_i = \eta^k \nabla_k \xi_i~.
\end{equation}

\noindent Here $\nabla_i$ denotes the covariant derivative, and round brackets denote symmetrization with respect to indices they embrace. All formulae that we use in this contribution are coordinate independent. However, for intuitive illustrations, it is convenient to introduce special coordinates that follow the time and axial symmetries and closely resemble the Newtonian spherical coordinates $[t, r, \phi, \theta]$ used in the previous section. In these coordinates 

\begin{equation}
\label{Killing-coordinates}
\eta^i = \delta^i_{~t}\,~~ \xi^i = \delta^i_{~\phi},
\end{equation}

\noindent $\delta^i_{~k}$ being the Kronecker delta, and the metric takes the form,

\begin{equation}
\label{metric}
ds^2 = (\eta \eta)\, dt^2 + 2 (\eta \xi)\, dt d\phi + (\xi \xi)\, d\phi^2 + g_{rr}\,dr^2 + g_{\theta \theta}\,d\theta^2,
\end{equation}

\noindent with a notation convention $x^iy^kg_{ik} \equiv (xy)$. The metric does not depend on $t$ and $\phi$, and we assume that it has an equatorial symmetry ``plane'', $\theta = \pi/2$. The geodesic circular motion is characterized by the four-velocity, $u^i = dx^i/ds$ that obeys,

\begin{equation}
\label{geodesic-equation}
u^k \nabla_k u^i = 0,~~u^i = A \left ( \eta^i + \Omega \xi^i \right ),~~ u^i u^k g_{ik} = 1.
\end{equation}

\noindent Here $\Omega$ is the angular velocity, and $A^{-2} = (\eta \eta) + 
2\Omega\,(\eta \xi) + \Omega^2\,(\xi \xi) > 0$. From (\ref{Killing}) and (\ref{geodesic-equation}) it follows that the energy defined by ${\cal E} = (\eta u) = u_t$, and the angular momentum defined by ${\cal L} = -(\xi u) = -u_{\phi}$, are constant of motion, because they obey

\begin{equation}
\label{constant-motion}
u^k \nabla_k {\cal E} = 0,~~u^k \nabla_k {\cal L} = 0.
\end{equation}

\noindent Obviously, the specific angular momentum defined by $\ell = {\cal L}/{\cal E}$, is also a constant of motion. The angular velocity $\Omega$ and the specific angular momentum $\ell$ are related by 

\begin{equation}
\label{velocity-momentum}
\Omega = - {{(\eta \eta)\ell + (\eta \xi)} \over {(\eta \xi)\ell + (\xi \xi)}},~~
\ell = - {{(\xi \xi)\Omega + (\eta \xi)} \over {(\eta \xi)\Omega + (\eta \eta)}}.
\end{equation}

\noindent The standard definition of the effective potential is,

\begin{equation}
\label{definition-effective-potential}
{\cal U}_{\rm eff} = -{1 \over 2} \ln \left ( g^{tt} - 2 \ell g^{t \phi} + \ell^2 g^{\phi \phi}\right ).
\end{equation}

\noindent This, together with the identity $1 = g^{ik} u_i u_k$ yields,

\begin{equation}
\label{unit-vector}
1 = {\cal E}^2\,e^{- 2{\cal U}_{\rm eff}}, ~~{\rm or}~~ 0 = \ln {\cal E} - {\cal U}_{\rm eff}.
\end{equation}

\noindent The three metric components that appear in the effective potential formula (\ref{definition-effective-potential}) may be invariantly defined in terms of the Killing vecors $\eta^i$ and $\xi^i$, and expressed by three scalar funtions: ${\tilde \Phi}$ being the gravitational potential, ${\tilde \omega}$ being the angular velocity of frame-dragging, and ${\tilde r}$ being the circumferential 
axial radius, 

\begin{subequation}
\label{metric-definition}
\begin{eqnarray}
g^{tt} &=& e^{- 2\,{\tilde \Phi}}\,\,\,\,\,\,\, \equiv {(\xi \xi) \over {(\eta \eta)(\xi \xi) - (\eta \xi)^2}},\\
g^{t\phi} &=& e^{- 2\,{\tilde \Phi}}{\tilde \omega}\,\,\, \equiv {{-(\eta \xi)} \over {(\eta \eta)(\xi \xi) - (\eta \xi)^2}},\\
- g^{\phi \phi} &=& e^{-2\,{\tilde \Phi}}{1 \over {\tilde r}^2} \equiv {{-(\eta \eta)} \over {(\eta \eta)(\xi \xi) - (\eta \xi)^2}}.
\end{eqnarray}
\end{subequation}

\noindent From these definitions one easily recovers the explicit expressions for the three scalars,

\begin{equation}
\label{three-scalars}
{\tilde \omega} = - {{(\eta \xi)} \over {(\xi \xi)}},~~
{\tilde r}^2 = - {{(\xi \xi)} \over {(\eta \eta)}} > 0,~~
e^{2\,\tilde \Phi} = (\eta \eta) + {\tilde \omega}(\eta \xi) > 0,
\end{equation}

\noindent which show that their Newtonian limits are, as they should be, ${\tilde \Phi} \rightarrow \Phi$, $\,{\tilde \omega} \rightarrow 0$, $\,{\tilde r} \rightarrow r \sin \theta$. From these limits, from (\ref{definition-effective-potential}), (\ref{metric-definition}), (\ref{three-scalars}), and from $\ell \rightarrow L\,$, $\ell^2/{\tilde r}^2 \ll 1$, one concludes that in the Newtonian limit, the relativistic effective potential goes to the Newtonian effective potential,

\begin{equation}
\label{another-effective-potential}
{\cal U}_{\rm eff} = 
{\tilde \Phi} - {1\over 2} \ln \left (1 - 2{\tilde \omega}\ell -
{\ell^2 \over {{\tilde r}^2}} \right ) \rightarrow U_{\rm eff}.
\end{equation}

\noindent Because ${\cal E} \rightarrow 1 + E$, and $E \ll 1$, one sees that equation (\ref{unit-vector}) has the correct Newtonian limit,
$\left \{0 = \ln{\cal E} - {\cal U}_{\rm eff}~\right\} \rightarrow \left\{~E = U_{\rm eff} ~\right\}$. 

%%%%%%%%%%%%%%%%%%%%%%%%%%%%%%%%%%%%%%%%%%%%%%%%%%%%%%%%%%%%%%%%%%%%
%%
\section{The four easy pieces}
\label{derivation-4}
%%
%%%%%%%%%%%%%%%%%%%%%%%%%%%%%%%%%%%%%%%%%%%%%%%%%%%%%%%%%%%%%%%%%%%%

We now derive simple and practical formulae for both radial and vertical orbital epicyclic frequencies in four lines of easy, fully transparent, algebra. The dot denotes derivative with respect to the proper time $s$, and $x$ denotes either radial $r$ or polar angle $\theta$ coordinate. We use the standard definitions for energy, specific angular momentum and effective potential, ${\cal E} = u_t,~\,\ell = - {u_{\phi}/u_t}\,$, ${\cal U}_{\rm eff} = -(1/2)\ln (g^{tt} - 2 \ell g^{t \phi} + \ell^2 g^{\phi \phi})$. They have been recalled and explained in the previous section. As in the Newtonian derivation described in section \ref{Newton-2}, we consider small oscillations with $\delta \ell \equiv 0$, $\,\delta {\cal E} \not = 0$, $\,\delta {\dot {\cal E}} = 0$, that occur either in $r$ or in $\theta$ direction, $x(s)-x_0 = \delta x$, $\,u^x \equiv dx/ds = \delta{\dot x}$. The first $r$ and $\theta$ derivatives of the effective potential are zero, which corresponds to an unperturbed circular orbit at the equatorial plane. The Taylor expansion starts from the second term, $\delta x^2$, and also ends there in the lowest order. Our four easy pieces consist of

\begin{subequation}
\begin{eqnarray}
\label{derivation}
1 &=& u_t u_t g^{tt} + 2 u_t u_{\phi} g^{t \phi} + u_{\phi} u_{\phi} g^{\phi \phi} + u^x u^x g_{xx},\\
{\cal E}^{-2} &=& e^{-2{\cal U}_{\rm eff}} + {g_{xx} \over {\cal E}^2}\left ( \delta {\dot x}\right )^2,\\
-2{{\delta {\cal E}}\over {{\cal E}^3_0}} 
&=& {1\over 2} \left( {\partial^2 \over \partial x^2}\,\,
 e^{-2{\cal U}_{\rm eff}}\right ) (\delta x)^2 + {g_{xx} \over {{\cal E}^2_0}} \left ( \delta {\dot x}\right )^2,\\
-2 {{\delta {\dot {\cal E}}} \over { {\cal E}^3_0}} 
&=& (\delta {\dot x})\left [ \left( {\partial^2 \over \partial x^2}\,\,
 e^{-2{\cal U}_{\rm eff}}\right ) \delta {x} + 2 {g_{xx} \over {{\cal E}^2_0}} \delta{\ddot  x} \right ].
\end{eqnarray}
\end{subequation}

\noindent The last line, with $\delta {\dot {\cal E}}=0,~\delta {\dot x} \not = 0$, has obviously the form of a simple har\-mo\-nic os\-cila\-tor equa\-tion, $0 = \omega^2_x \delta x + \delta {\ddot x}$ and, with the zeroth order version of eq. (26b), this yields the final result, 

\begin{equation}
\label{final-result}
\omega^2_x
 = \left ({{\partial^2 {\cal U}_{\rm eff}} \over {\partial X^2}} \right ),
\end{equation}

\noindent with $dX^2 = - g_{xx} dx^2>0$ being the proper length in the $x$ direction.

We see that exactly as in Newton's theory, both radial and vertical epicyclic frequencies (squared) are equal to second derivatives of the invariantly defined effective potential, with respect to the invariantly defined coordinates in radial and vertical directions.

%%%%%%%%%%%%%%%%%%%%%%%%%%%%%%%%%%%%%%%%%%%%%%%%%%%%%%%%%%%%%%%%%%%%
%%
\section{Practical calculations}
\label{examples-5}
%%
%%%%%%%%%%%%%%%%%%%%%%%%%%%%%%%%%%%%%%%%%%%%%%%%%%%%%%%%%%%%%%%%%%%%

We gave a simple derivation of general, physically clear, formulae (\ref{final-result}) for the two epicyclic oscillation frequencies. In this section we show that they are also simple to use in practical calculations, in particular when the metric is numerically constructed. Indeed, in order to solve second order Einstein's field equations and numerically construct a spacetime (e.g., outside a rotating neutron star), one must calculate the metric, its first and second derivatives. Thus, one stores during the calculations,

\begin{subequation}
\label{metric-known}
\begin{eqnarray}
&~&g^{tt},~~g^{t \phi},~~g^{\phi \phi},~~g_{rr},~~g_{\theta \theta},\\
&~&g^{tt}_{[r]},~~g^{t\phi}_{[r]},~~g^{\phi \phi}_{[r]},~~
g^{tt}_{[\theta]},~~g^{t\phi}_{[\theta]},~~g^{\phi \phi}_{[\theta]},\\
&~&g^{tt}_{[rr]},~~g^{t\phi}_{[rr]},~~g^{\phi \phi}_{[rr]},~~
g^{tt}_{[\theta \theta]},~~g^{t\phi}_{[\theta \theta]},~~g^{\phi \phi}_{[\theta \theta]},
\end{eqnarray}
\end{subequation}

\noindent where derivatives of a quantity $Y$ are indicated by indices in square brackets, $\partial Y/\partial x \equiv Y_{[x]}$, $\partial^2 Y/\partial x^2 \equiv Y_{[xx]}$. Let us now restrict to the equatorial symmetry plane, where all 
$\partial g^{ik}/\partial \theta \equiv g^{ik}_{[\theta]} = 0$. The condition for the circular orbit, $\partial {\cal U}_{\rm eff}/\partial x = 0$ is thus trivially fulfilled for $x = \theta$, and for $x =r$ it gives the equation for the Keplerian angular mometum for circular orbits,

\begin{equation}
\label{Einstein-Kepler-equation}
g^{tt}_{[r]} - 2\,\ell\,g^{t\phi}_{[r]} + \ell^2\,g^{\phi \phi}_{[r]} =0
\end{equation}

\noindent which has the solution,

\begin{equation}
\label{Einstein-Kepler-solution}
\ell_K = {{g^{t \phi}_{[r]} \pm \left [ (g^{t\phi}_{[r]})^2 - g^{\phi \phi}_{[r]}\,g^{tt}_{[r]} \right ]^{1\over 2}} \over {g^{\phi \phi}_{[r]}}},
\end{equation}

\noindent with the plus sign before the square bracket holding for $\ell_K > 0$ and the minus sign for $\ell_K < 0$. 

We use (\ref{velocity-momentum}), and the radial derivative of eq. (26a)
for circular orbits, to give two expressions for the orbital Keplerian angular velocity in terms of the first derivatives of the metric,

\begin{equation}
\label{Einstein-Kepler-Omega}
\Omega_K = - {{g_{tt}\ell_K + g_{t \phi}} \over {g_{t \phi} \ell_K + g_{\phi \phi}}}
= 
{{-(g_{t \phi})_{[r]} \pm \left [ (g_{t\phi})_{[r]}^2 - (g_{\phi \phi})_{[r]}\,(g_{t t})_{[r]} \right ]^{1\over 2}} \over {(g_{\phi \phi})_{[r]}}}.
\end{equation}

Our general result (\ref{final-result}), explicitly expressed in terms of the metric and its first and second derivatives, takes the form,

\begin{subequation}
\label{final-result-explicit}
\begin{eqnarray}
\omega^2_r &=&
+ \,{1 \over {2 g_{rr}}}\,\,
{{g^{tt}_{[rr]} - 2\ell_K \,g^{t\phi}_{[rr]} + \ell^2_K \,g^{\phi \phi}_{[rr]}} 
\over
{g^{tt} - 2\ell_K\, g^{t \phi} + \ell^2_K\, g^{\phi \phi}}},\\
\omega^2_{\theta} &=&
+ \,{1\over {2 g_{\theta \theta}}}\,\,
{{g^{tt}_{[\theta \theta]} - 2\ell_K\, g^{t\phi}_{[\theta \theta]} + \ell^2_K\, g^{\phi \phi}_{[\theta \theta]}} \over
{g^{tt} - 2\ell_K\, g^{t \phi} + \ell^2_K\, g^{\phi \phi}}}.
\end{eqnarray}
\end{subequation}

\noindent These frequencies $\omega_r$, $\,\omega_{\theta}$ are measured with respect to the proper time of a comoving observer. After dividing by the squared redshift factor $(u^t)^2= e^{2{\cal U}_{\rm eff}}\left ( g_{tt} + \Omega_K \,g_{t \phi}\right)^{-2}$, one gets the ``observed'' frequencies at ``infinity'', i.e., those measured with respect to the time $t$ of an astronomer at rest very far from the source:

\begin{subequation}
\label{final-explicit-observer}
\begin{eqnarray}
\Omega^2_r &=&
 \,{{\left ( g_{tt} + \Omega_K \,g_{t \phi}\right)^2} \over {2 g_{rr}}}\,\,
\left [ g^{tt}_{[rr]} - 2\ell_K \,g^{t\phi}_{[rr]} + \ell^2_K \,g^{\phi \phi}_{[rr]}\right ],\\
\Omega^2_{\theta} &=&
 \,{{\left (g_{tt} + \Omega_K \,g_{t \phi}\right)^2} \over {2 g_{\theta \theta}}}\,\,
\left [ g^{tt}_{[\theta \theta]} - 2\ell_K\, g^{t\phi}_{[\theta \theta]} + \ell^2_K\, g^{\phi \phi}_{[\theta \theta]} \right].
\end{eqnarray}
\end{subequation}

\noindent For an aplication of these formulae to realistic models of rotating
neutron stars, see e.g., \inlinecite{Klu2004}.

By expanding $(\,g^{tt}_{[r]} - 2\ell_K \,g^{t\phi}_{[r]} + \ell^2_K \,g^{\phi \phi}_{[r]}\,)_{[r]} = 0$, one arrives at

\begin{equation}
\label{stability-final}
\Omega^2_r = \pm\,{\cal A}\,\left ( {d\ell_K \over dr}\right )\,
\left [ (g^{t\phi}_{[r]})^2 - g^{\phi \phi}_{[r]}\,g^{tt}_{[r]} \right ]^{1\over 2},
~~{\cal A} \equiv {{\left ( g_{tt} + \Omega_K \,g_{t \phi}\right)^2} \over {(- g_{rr})}} > 0.
\end{equation}

\noindent Because by construction (\ref{Einstein-Kepler-solution}), $\pm\,{\rm sgn}\, \ell_K = [ (g^{t\phi}_{[r]})^2 - g^{\phi \phi}_{[r]}\,g^{tt}_{[r]} ]^{1/2}>0$, one concludes that $\,\Omega^2_r > 0$ if and only if $\,d |\ell_K|/dr > 0$. 

Thus, the relativistic Rayleigh criterion for circular Keplerian orbits is the same as the Newtonian one [cf. equation (\ref{Newton-Rayleigh})]: for stability, the absolute value of specific angular momentum should be an increasing function of the orbital radius. And, just as in Newtonian theory, the epicyclic frequencies can be derived from the second derivatives of an effective potential (see eq. [27]). 

\begin{acknowledgements}
We thank Nordita for generous hospitality and support. The calculations reported here have been done, and the article written, during our several visits to Nordita, that Nordita supported in full. We also thank Dr. Nikolaos Stergioulas
for implementing our formulae numerically. 
\end{acknowledgements}

\end{article}

\begin{thebibliography}

%%%%%%%%%%%%%%%%%%%%%%%%%%%%%%%%%%%%%%%%%%%%%%%%%%%%%%%%%%%%%%
\bibitem[\protect\citeauthoryear{Abramowicz \& Klu{\'z}niak}{2000}]
{AbraKlu2000}
Abramowicz, M.A., Klu{\'z}niak, W.: 2000,
unpublished research notes,
%%%%%%%%%%%%%%%%%%%%%%%%%%%%%%%%%%%%%%%%%%%%%%%%%%%%%%%%%%%%%%

%%%%%%%%%%%%%%%%%%%%%%%%%%%%%%%%%%%%%%%%%%%%%%%%%%%%%%%%%%%%%%
\bibitem[\protect\citeauthoryear{Abramowicz \& al.}{2003}]{AbrAlmKluTha2003}
Abramowicz, M.A., Almergren, G.J.E., Klu{\'z}niak, W., Thampan, A.V.: 2003
\\gr-qc/0312070
%%%%%%%%%%%%%%%%%%%%%%%%%%%%%%%%%%%%%%%%%%%%%%%%%%%%%%%%%%%%%%

%%%%%%%%%%%%%%%%%%%%%%%%%%%%%%%%%%%%%%%%%%%%%%%%%%%%%%%%%%%%%%
\bibitem[\protect\citeauthoryear{Amsterdamski \& al.}{2002}] 
{AmsEtAL2002}  
Amsterdamski, P., Bulik, T., Gondek-Rosi{\'n}ska, D., Klu{\'z}niak, W.: 2002,
{\it A\&A}, {\bf 381}, L21
%%%%%%%%%%%%%%%%%%%%%%%%%%%%%%%%%%%%%%%%%%%%%%%%%%%%%%%%%%%%%%

%%%%%%%%%%%%%%%%%%%%%%%%%%%%%%%%%%%%%%%%%%%%%%%%%%%%%%%%%%%%%%
\bibitem[\protect\citeauthoryear{Klu\'zniak \& al.}{2004}]{Klu2004}
Klu\'zniak, W., Abramowicz, M. A., Kato, S., Lee, W. H., Stergioulas, N.:
2004, {\it ApJ}, {\bf 603}, L89	
%%%%%%%%%%%%%%%%%%%%%%%%%%%%%%%%%%%%%%%%%%%%%%%%%%%%%%%%%%%%%%

%%%%%%%%%%%%%%%%%%%%%%%%%%%%%%%%%%%%%%%%%%%%%%%%%%%%%%%%%%%%%%
\bibitem[\protect\citeauthoryear{Klu\'zniak \& al.}{2001}]{KlEtAL2001}
Klu{\'z}niak, W., Bulik, T., Gondek-Rosi{\'n}ska, D.,: 2001,
in {\it  Exploring the gamma-ray universe. Proceedings of the Fourth INTEGRAL Workshop, 4-8 September 2000, Alicante, Spain.} Editor: B. Battrick, Scientific editors: A. Gimenez, V. Reglero \& C. Winkler. ESA SP-459, Noordwijk: ESA Publications Division, pp. 301-304
%%%%%%%%%%%%%%%%%%%%%%%%%%%%%%%%%%%%%%%%%%%%%%%%%%%%%%%%%%%%%%

%%%%%%%%%%%%%%%%%%%%%%%%%%%%%%%%%%%%%%%%%%%%%%%%%%%%%%%%%%%%%%
\bibitem[\protect\citeauthoryear{Markovi{\'c}}{2000}]{Mar2000} 
Markovi{\'c}, D.: 2000, astro-ph/0009450
%%%%%%%%%%%%%%%%%%%%%%%%%%%%%%%%%%%%%%%%%%%%%%%%%%%%%%%%%%%%%%

%%%%%%%%%%%%%%%%%%%%%%%%%%%%%%%%%%%%%%%%%%%%%%%%%%%%%%%%%%%%%%
\bibitem[\protect\citeauthoryear{Nowak \& Lehr}{1999}]{NowLeh1999} 
Nowak, M., Lehr. D.: 1999,
in {\it Theory of Black Hole Accretion Disks},  
eds. Abra\-mo\-wicz M.A., Bj{\"o}rnsson G., Pringle J.E., 
Cambridge University Press, Cambridge
%%%%%%%%%%%%%%%%%%%%%%%%%%%%%%%%%%%%%%%%%%%%%%%%%%%%%%%%%%%%%%

%%%%%%%%%%%%%%%%%%%%%%%%%%%%%%%%%%%%%%%%%%%%%%%%%%%%%%%%%%%%%%
\bibitem[\protect\citeauthoryear{Okazaki, 
Kato \& Fukue}{1987}]{OkaKatFuk1987} Okazaki, A. T., 
Kato, S., Fukue, J.: 1987,
{\it Publ. Astron. Soc. Japan}, {\bf 39}, 457
%%%%%%%%%%%%%%%%%%%%%%%%%%%%%%%%%%%%%%%%%%%%%%%%%%%%%%%%%%%%%%

%%%%%%%%%%%%%%%%%%%%%%%%%%%%%%%%%%%%%%%%%%%%%%%%%%%%%%%%%%%%%%
\bibitem[\protect\citeauthoryear{Perez, Silbergleit, Wagoner
 \& Lehr}{1997}]{Perez97}
 Perez, C. A., Silbergleit, A. S., Wagoner, R. V., Lehr, D. E.:1997,
{\it ApJ}, {\bf 476}, 589
%%%%%%%%%%%%%%%%%%%%%%%%%%%%%%%%%%%%%%%%%%%%%%%%%%%%%%%%%%%%%%

%%%%%%%%%%%%%%%%%%%%%%%%%%%%%%%%%%%%%%%%%%%%%%%%%%%%%%%%%%%%%%
\bibitem[\protect\citeauthoryear{Wald}{1984}]{Wal1984}
Wald, R.W.: 1984,
{\it General Relativity}, University of Chicago Press, Chicago
%%%%%%%%%%%%%%%%%%%%%%%%%%%%%%%%%%%%%%%%%%%%%%%%%%%%%%%%%%%%%%

\end{thebibliography}
\end{document}